\documentclass[showpacs,amsmath,amssymb,aps,showkeys,floatfix,prd,a4paper]{revtex4}

\usepackage[dvips]{graphicx}
\usepackage{dcolumn}
\usepackage{bm}
\usepackage{epsfig}
\usepackage{amsfonts}
\usepackage{amssymb,amscd}

\def\lsim{\raise0.3ex\hbox{$<$\kern-0.75em\raise-1.1ex\hbox{$\sim$}}}

\def\gsim{\raise0.3ex\hbox{$>$\kern-0.75em\raise-1.1ex\hbox{$\sim$}}}

\newcommand{\be}{\begin{equation}}

\newcommand{\ee}{\end{equation}}

\def\beq{\begin{equation}}

\def\eeq{\end{equation}}

\def\beqa{\begin{eqnarray}}

\def\eeqa{\end{eqnarray}}

\newcommand{\rd}{\mbox{\boldmath $\Delta$}}

\newcommand{\ba}{\begin{eqnarray}}

\newcommand{\rr}{\mbox{\boldmath $r$}}

\newcommand{\rb}{\mbox{\boldmath $b$}}

\def\gappeq{\mathrel{\rlap {\raise.5ex\hbox{$>$}}

{\lower.5ex\hbox{$\sim$}}}}

\def\lappeq{\mathrel{\rlap{\raise.5ex\hbox{$<$}}

{\lower.5ex\hbox{$\sim$}}}}

\def\Toprel#1\over#2{\mathrel{\mathop{#2}\limits^{#1}}}

\begin{document}

\begin{flushright}
LU TP 16-XX\\
May 2016
\vskip1cm
\end{flushright}

\title{Exclusive vector meson production with a leading neutron in  photon - hadron interactions at hadronic colliders}
\author{V.P. Gon\c{c}alves $^{1,2}$,  B.D.  Moreira$^{3}$,  F.S. Navarra$^3$ and D. Spiering $^{3}$}
\affiliation{$^1$ Department of Astronomy and Theoretical Physics, Lund University, SE-223 62 Lund, Sweden \\  
$^{2}$ High and Medium Energy Group, Instituto de F\'{\i}sica e Matem\'atica,  Universidade Federal de Pelotas\\
Caixa Postal 354,  96010-900, Pelotas, RS, Brazil.\\
$^3$Instituto de F\'{\i}sica, Universidade de S\~{a}o Paulo,
C.P. 66318,  05315-970 S\~{a}o Paulo, SP, Brazil\\
}

\begin{abstract}
In this paper we  study leading neutron production in  photon - hadron interactions which take place in $pp$ and $pA$ 
collisions at large impact parameters. Using a model that describes the recent leading neutron data at HERA, we consider  
exclusive vector meson production in association with a leading neutron in $pp/pA$ collisions at RHIC and LHC energies. 
The total cross sections and rapidity distributions of  $\rho$, $\phi$ and $J/\Psi$ produced together with a leading neutron 
are computed. Our results indicate that the study of these processes is feasible and that it can be used to improve the 
understanding of  leading neutron processes and of exclusive vector meson production. 

\end{abstract}



\maketitle

\section{Introduction}

Understanding leading particle production is crucial to understand  forward physics at hadron colliders 
and also cosmic ray physics \cite{FP}. Indeed, the interpretation of cosmic ray  data is strongly 
dependent on the accurate knowledge of the leading baryon momentum spectrum  and its energy dependence 
(See e.g. Ref. \cite{cr}). Moreover,  particle production at forward rapidities and high energies probes 
the QCD dynamics at very small - $x$, where non-linear effects associated to high gluonic density in the 
target are expected to contribute significantly \cite{hdqcd}. This new regime of the QCD dynamics is a 
field of intense activity and the exclusive production of vector mesons in $ep (A)$ collisions and in 
ultraperipheral hadronic collisions  is one of the most promising observables to constrain the main 
properties of the theory (See, e.g. Refs.  \cite{kmw,vicmag_mesons1,vicmag_update,amir_armesto,bruno1,
bruno2,diego_rho}).

Leading neutron production has been investigated in $ep$ collisions at HERA, from where  we have  high 
precision experimental data on semi - inclusive  $e + p \rightarrow e + n + X$ processes \cite{lpdata2}  
as well as on  exclusive $\rho$ photoproduction associated with a leading neutron   
($\gamma p \rightarrow \rho^0 \pi^+ n$) \cite{rhoLN_HERA}. In these processes the  incident proton is 
converted into a neutron via pion emission. In Refs. \cite{nosLN,nosLN2} we proposed an unified  
description of  inclusive and exclusive processes with a leading neutron, based on the color dipole 
formalism, and we have demonstrated that the available experimental HERA data on  the  $x_L$ 
(Feynman momentum) distribution of leading neutrons can be very well described in this approach. An 
important property of our approach is that its main elements are constrained by the  HERA data on 
processes without a leading neutron. As a consequence, our analysis of  leading neutron data  
has put limits on the magnitude of the nonperturbative absorptive corrections and on the models of the 
pion flux (which describes the pion emission by the incident proton). Moreover, we were able  to present 
parameter - free predictions for the inclusive and exclusive processes with a leading neutron at the 
energies of the future $ep$ colliders \cite{Boer,Accardi,LHeC}. Unfortunately, in view of the 
construction schedule of the these new colliders,  these predictions will 
only be tested in a distant future. Given the impact of  leading neutron production in forward physics, 
it is fundamental to consider alternative ways to learn more about this process (See. e.g. 
Refs. \cite{kopeliovich,Ryutin}). 

In this paper we propose the study of the  leading neutron production in the photon - hadron ($\gamma h$) 
interactions, which are present  in hadronic collisions \cite{upc}. In particular, we 
will consider  exclusive vector meson production associated with a leading neutron in $\gamma p$ 
interactions at $pp$ and $pA$ collisions. Recent theoretical  and experimental studies have demonstrated  
that  hadronic colliders can also be used to study photon - hadron and photon - photon interactions in 
a new kinematical range and that several open questions in the theory of strong interactions can be 
investigated by  analysing  different final states produced in these reactions (For a recent review 
see Ref. \cite{FP}). As we will demonstrate below, such conclusion  is also valid for leading neutron 
processes. In what follows we will investigate  the exclusive $\rho$, $\phi$ and $J/\Psi$ production 
associated with a leading neutron in $pp$ and $pA$ collisions at RHIC and LHC energies and present our 
estimates for the total cross section and rapidity distributions of these distinct final states.  Our 
goal is to demonstrate that the experimental analysis of these processes is feasible and that they may 
be used to study  leading neutron physics  as well as to study exclusive vector meson 
production.

This paper is organized as follows. In the next Section we present the main concepts in  photon - induced 
interactions and discuss exclusive vector meson production associated with a leading neutron. In Section 
\ref{res} we present our predictions for the rapidity distributions and total cross sections for  
exclusive $\rho$, $\phi$ and $J/\Psi$ production associated with a leading neutron in $pp$ and $pA$ 
collisions at RHIC and LHC energies. Finally, in Section \ref{conc} we summarize our main conclusions.

\begin{figure}[t]
\begin{center}
\scalebox{0.45}{\includegraphics{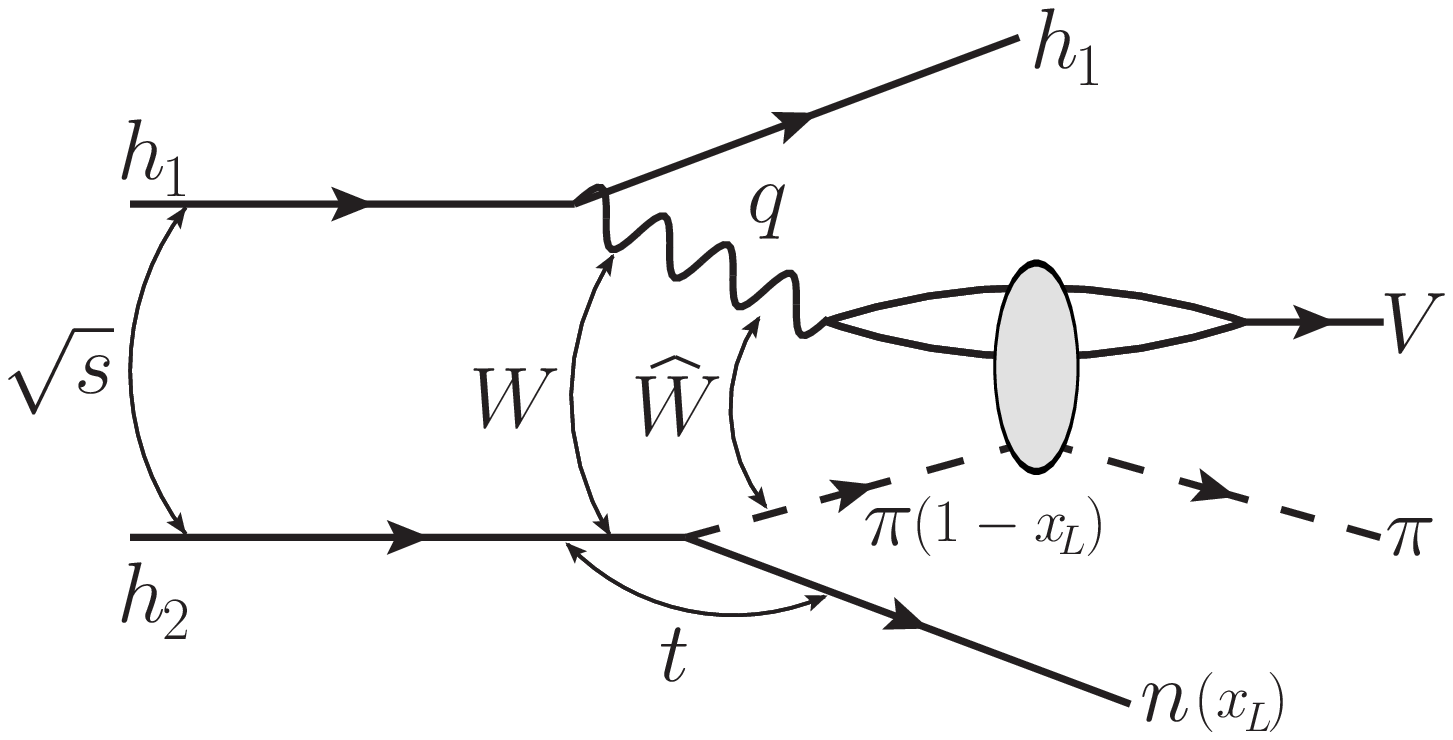}}
\scalebox{0.45}{\includegraphics{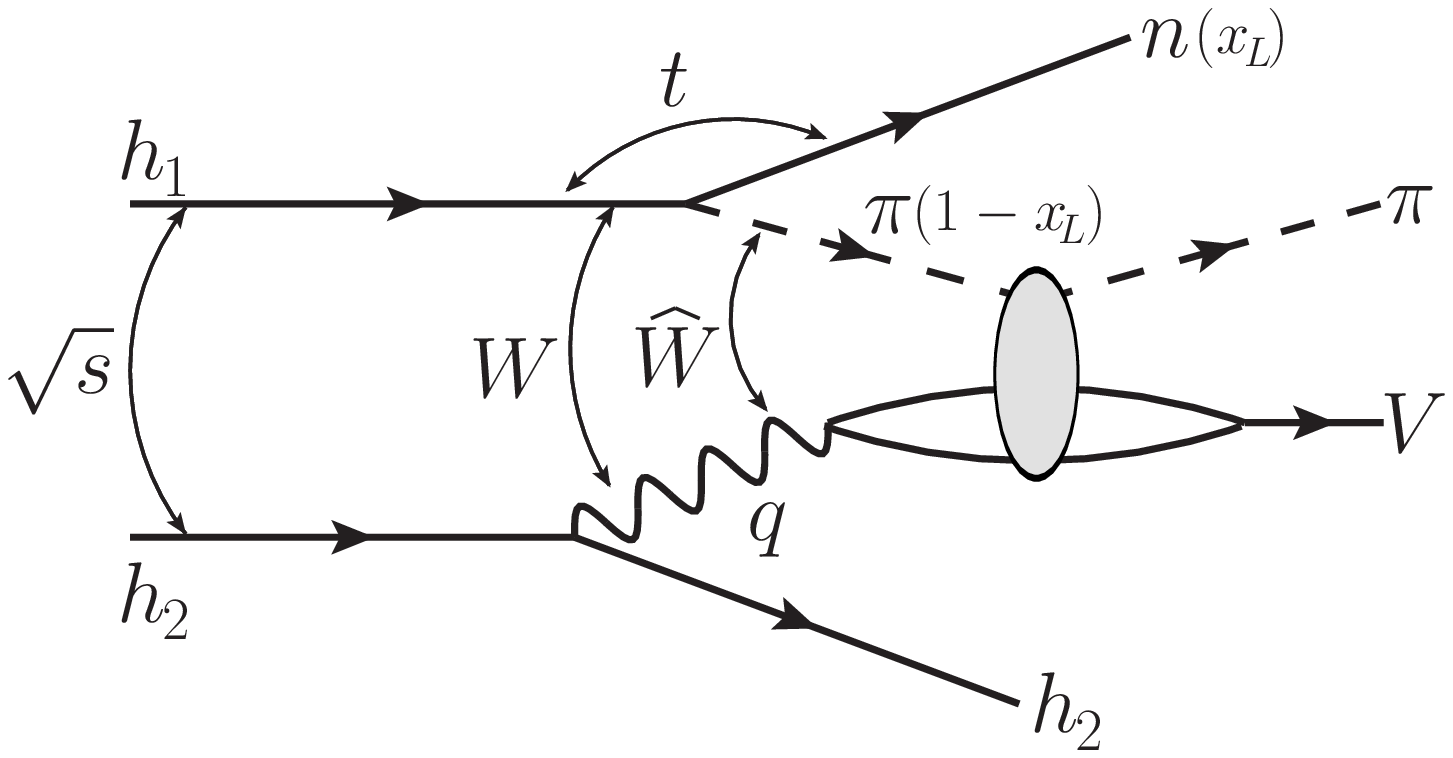}}
\caption{Exclusive vector meson production associated with a leading neutron in $\gamma h$ 
interactions at hadronic colliders.}
\label{diagramas}
\end{center}
\end{figure}

\section{Photon - induced interactions and  exclusive vector meson production associated with a 
leading neutron}

In this section we will present a brief review of the formalism needed to describe the vector meson 
production associated with a leading neutron in photon - induced interactions at hadronic collisions. 
We refer the reader to our previous papers \cite{bruno1,bruno2,nosLN,nosLN2} for a more detailed 
discussion. At high energies, the incident charged hadrons  (proton or nuclei)  generate strong 
electromagnetic fields, which can be represented in terms of an equivalent photon flux. As a consequence, 
in a hadronic collision, a photon stemming from the electromagnetic field of one of the two colliding 
hadrons can interact with one photon coming from the other hadron (photon - photon process) or it can 
interact directly with the other hadron (photon - hadron process) \cite{upc,epa}. In this paper we will 
focus on  the latter. A basic property of  these photon - induced interactions is that 
the cross section can be factorized in terms of the  equivalent flux of photons (from the hadron 
projectile) and the  photon - target cross section, with the photon flux being well known. Consequently,  
the cross section of a photon - induced process in hadronic collisions is a direct probe of the photon - 
hadron cross section. Therefore, the study of  these processes in hadronic colliders can be considered 
complementary to the analysis performed at HERA, as demonstrated by recent results on exclusive vector 
meson photoproduction in $pp/pA$ collisions.  The main difference in comparison to HERA and future $ep$ 
colliders, is that photon - induced interactions are characterized by real photons, while in $ep$ 
colliders it  is possible to investigate  vector meson production at  different photon virtualities.   
We will extend the formalism used to treat photon - induced interactions to the description of the leading 
neutron processes.  Exclusive vector meson production associated with a leading neutron  
will be represented by the  diagrams in  Fig. \ref{diagramas}, which can be seen as a sequence of four 
factorizable subprocesses:  i) a photon is emitted by one of the incident hadrons; ii) this photon 
fluctuates into a quark-antiquark pair (the color dipole), iii) the color dipole interacts diffractively 
with the pion, while the  proton remains intact, and iv) the vector meson and the leading neutron are 
formed. The final state will be characterized by two rapidity gaps, one associated to the photon exchange  
and another to the colorless system exchanged  between the dipole and the pion, as well as by a pion 
produced at large rapidities  and by a leading neutron. The corresponding rapidity distribution will be  
given by 
\begin{eqnarray}
\frac{d\sigma \,\left[h_1 + h_2 \rightarrow   h_3 \otimes V  \otimes \pi + n \right]}{dY} = \left[\omega 
\frac{dN}{d\omega}|_{h_1}\,\sigma_{\gamma h_2 \rightarrow V  \otimes \pi + n}\left(\omega \right)\right]_{\omega_L} + 
\left[\omega \frac{dN}{d\omega}|_{h_2}\,\sigma_{\gamma h_1 \rightarrow V  \otimes \pi + n}
\left(\omega \right)\right]_{\omega_R}\,
\label{dsigdy}
\end{eqnarray}
where $h_3$ corresponds to the initial hadron ($h_1$ or $h_2$) which has emitted the photon, $Y$ is the rapidity 
of the vector meson $V$ ($ = \rho, \phi,J/\Psi$) in the final state, which can be determined by the photon 
energy $\omega$ in the collider frame and by mass $M_{V}$ of the vector meson [$Y\propto \ln \, ( \omega/M_{V})$].
The symbol $\otimes$ represents  a rapidity gap in the final state and $\omega_L \, (\propto e^{-Y})$ 
and $\omega_R \, (\propto e^{Y})$ denote photons from the $h_1$ and $h_2$ hadrons, respectively.  
In our study we will assume that the  equivalent photon spectrum $\frac{dN}{d\omega}$ of a relativistic proton 
is given by  \cite{Drees},
\begin{eqnarray}
\frac{dN_{\gamma/p}(\omega)}{d\omega} =  \frac{\alpha_{\mathrm{em}}}{2 \pi\, \omega} \left[ 1 + \left(1 -
\frac{2\,\omega}{\sqrt{s}}\right)^2 \right] 
\left( \ln{\Omega} - \frac{11}{6} + \frac{3}{\Omega}  - \frac{3}{2 \,\Omega^2} + \frac{1}{3 \,\Omega^3} \right) \,,
\label{eq:photon_spectrum}
\end{eqnarray}
with the notation $\Omega = 1 + [\,(0.71 \,\mathrm{GeV}^2)/Q_{\mathrm{min}}^2\,]$, 
$Q_{\mathrm{min}}^2= \omega^2/[\,\gamma_L^2 \,(1-2\,\omega /\sqrt{s})\,] \approx 
(\omega/\gamma_L)^2$, $\gamma_L$ is the Lorentz boost  of a single beam and $\sqrt{s}$ is  the c.m.s energy of the
hadron-hadron system. The equivalent photon flux of a nucleus is  given by \cite{upc}
\begin{eqnarray}
\frac{dN_{\gamma/A}\,(\omega)}{d\omega}= \frac{2\,Z^2\alpha_{em}}{\pi\,\omega}\, \left[\bar{\eta}\,K_0\,(\bar{\eta})\, 
K_1\,(\bar{\eta})+ \frac{\bar{\eta}^2}{2}\,{\cal{U}}(\bar{\eta}) \right]\,
\label{fluxint}
\end{eqnarray}
where   $\bar{\eta}=\omega\,(R_{h_1} + R_{h_2})/\gamma_L$, $K_{0,1}$ are the modified Bessel functions of second kind and  
${\cal{U}}(\bar{\eta}) = K_1^2\,(\bar{\eta})-  K_0^2\,(\bar{\eta})$. One of the main differences between the photon fluxes 
for the proton and for the nucleus is that the latter is enhanced by a factor $Z^2$. This implies that the rapidity distribution 
of the vector mesons produced in $pA$ collisions will be asymmetric and determined by $\gamma p$ interactions, with the photon 
coming from the nucleus. In contrast, the rapidity distribution for $pp$ collisions will be symmetric with respect to $Y=0$. 

Following Refs. \cite{nosLN,nosLN2} we will assume that the absorptive corrections associated to soft rescatterings \cite{speth,pirner} 
can be approximated by a constant factor $\cal{K}$, which implies that the total cross section for the process 
$\gamma p \rightarrow V  \otimes \pi + n$ can be expressed by
\beq
\sigma_{\gamma p \rightarrow V  \otimes \pi + n} (W^2) = {\cal{K}} \cdot \int dx_L dt \, f_{\pi/p} (x_L,t) \cdot 
\sigma_{\gamma \pi \rightarrow V  \otimes \pi}(\hat{W}^2)
\label{crossgen}
\eeq
where  ${W}$ is the center-of-mass energy of the  photon-proton system, $x_L$ is the proton momentum fraction carried by the 
neutron and $t$ is the square of the four-momentum of the exchanged pion. Moreover,  $f_{\pi/p}$ is the flux of virtual pions emitted by 
the proton  and  $\sigma_{\gamma \pi \rightarrow V  \otimes \pi}(\hat{W}^2)$  is the cross section of the interaction between the  photon 
and the pion  at center-of-mass energy $\hat{W}$, which is given by  $\hat{W}^2 = (1-x_L) \, W^2$. As discussed in detail in Refs. 
\cite{nosLN,nosLN2}, the precise form of the pion flux  is still a subject under investigation. In what follows we will assume that the pion 
flux is given by
\beq
f_{\pi/p} (x_L,t)  = \frac{1}{4 \pi} \frac{2 g_{p \pi p}^2}{4  \pi} \frac{-t}{(t-m_{\pi}^2)^2} (1-x_L)^{1-2 \alpha(t)}  
[F(x_L,t)]^2
\label{genflux}
\eeq 
where $g_{p \pi p}^2/(4 \pi) = 14.4$ is the $ \pi^ 0 p p $ coupling constant, $m_{\pi}$ is the pion mass and $\alpha(t)$  is the Regge trajectory 
of the pion. In particular, we will consider that the form factor $F(x_L,t)$, which  accounts for the finite size of the nucleon and of the pion, 
can be expressed as follows \cite{kope}
\beq
F(x_L,t) =  \exp \left[ b (t-m_{\pi}^2) \right] \,\,\,\, , \,\,\,\, \alpha(t) = \alpha(t)_{\pi} 
\label{form3}
\eeq
where $\alpha_{\pi}(t) \simeq t$ (with $t$ in GeV$^2$)  and $b = 0.3$ GeV$^{-2}$. As demonstrated in Ref. \cite{nosLN2}, this model (denoted $f_3$ 
in \cite{nosLN2}) reproduces well the  HERA data on the exclusive $\rho$ photoproduction associated with a leading neutron.

In order to describe the $\gamma \pi \rightarrow V  \otimes \pi$ cross section, in Ref. \cite{nosLN2} we  extended the color dipole formalism 
(widely used in  exclusive $\gamma p$ processes) to $\gamma \pi$ interactions. Consequently, we assumed that  
$\sigma_{\gamma \pi \rightarrow V  \otimes \pi}$ can be expressed by 
\begin{eqnarray}
\sigma (\gamma \pi \rightarrow V \otimes \pi) =  \int_{-\infty}^0 \frac{d\sigma}{d\hat{t}}\, d\hat{t}  
= \frac{1}{16\pi}  \int_{-\infty}^0 |{\cal{A}}^{\gamma \pi \rightarrow V \pi }(x,\Delta)|^2 \, d\hat{t}\,\,,
\label{sctotal_intt}
\end{eqnarray}
with the scattering amplitude being given by 
\begin{eqnarray}
 {\cal A}^{\gamma \pi \rightarrow V \pi}(\hat{x},\Delta)  =  i
\int dz \, d^2\rr \, d^2\rb  e^{-i[\rb-(1-z)\rr].\rd} 
 \,\, (\Psi^{V*}\Psi)  \,\,2 {\cal{N}}_\pi(\hat{x},\rr,\rb)
\label{sigmatot2}
\end{eqnarray}
where $(\Psi^{V*}\Psi)$ denotes the overlap between the real photon and exclusive final state wave functions, which we assume to be given by the Gauss-LC 
model described in Ref. \cite{nosLN}. The variable  $z$ $(1-z)$ is the longitudinal momentum fraction of the quark (antiquark) and  $\Delta$ denotes the 
transverse momentum lost by the outgoing pion ($\hat{t} = - \Delta^2$). The variable $\rb$ is the transverse distance from the center of the target to the 
center of mass of the $q \bar{q}$  dipole and the factor  in the exponential  arises when one takes into account non-forward corrections to the wave functions 
\cite{non}. Moreover, $\mathcal{N}^\pi(\hat{x},\rr,\rb)$ is  the imaginary part of the forward amplitude of the scattering between a small dipole
(a colorless quark-antiquark pair) and a pion, at a given rapidity interval $y=\ln(1/\hat{x})$. This quantity is directly related to the QCD dynamics at high 
energies \cite{hdqcd}. As in Refs. \cite{nosLN,nosLN2} we will assume that $\mathcal{N}^\pi$  can be expressed in terms of the dipole-proton scattering amplitude  
$\mathcal{N}^p$, usually probed in the  inclusive and exclusive processes at HERA, as follows 
\begin{equation}
{\cal N}^\pi (\hat{x}, \rr, \rb) = R_q \cdot {\cal N}^p (\hat{x}, \rr, \rb) 
\label{doister}
\end{equation}
with $R_q$ being a constant. Moreover, we  will assume that   ${\cal{N}}_p (\hat{x},\rr,\rb)$ is given by the bCGC model proposed in Ref. \cite{kmw}:
\begin{eqnarray}
\mathcal{N}_p(\hat{x},\rr,{\rb}) =   
\left\{ \begin{array}{ll} 
{\mathcal N}_0\, \left(\frac{ r \, Q_{s,p}}{2}\right)^{2\left(\gamma_s + 
\frac{\ln (2/r Q_{s,p})}{\kappa \,\lambda \,y}\right)}  & \mbox{$r Q_{s,p} \le 2$} \\
 1 - \exp \left[-A\,\ln^2\,(B \, r \, Q_{s,p})\right]   & \mbox{$r Q_{s,p}  > 2$} 
\end{array} \right.
\label{eq:bcgc}
\end{eqnarray}
with  $y=\ln(1/\hat{x})$ and $\kappa = \chi''(\gamma_s)/\chi'(\gamma_s)$, where $\chi$ is the 
LO BFKL characteristic function \cite{bfkl}.  The coefficients $A$ and $B$  
are determined uniquely from the condition that $\mathcal{N}_p(\hat{x},\rr,\rb)$, and its derivative 
with respect to $rQ_s$, are continuous at $rQ_s=2$. 
In this model, the proton saturation scale $Q_{s,p}$ depends on the impact parameter:
\begin{equation} 
  Q_{s,p}\equiv Q_{s,p}(\hat{x},{\rb})=\left(\frac{x_0}{\hat{x}}\right)^{\frac{\lambda}{2}}\;
\left[\exp\left(-\frac{{b}^2}{2B_{\rm CGC}}\right)\right]^{\frac{1}{2\gamma_s}}.
\label{newqs}
\end{equation}
Following \cite{amir} we will assume in what follows that  $\gamma_s = 0.6599$, $B_{CGC} = 5.5$ GeV$^{-2}$,
$\mathcal{N}_0 = 0.3358$, $x_0 = 0.00105 \times 10^{-5}$ and $\lambda = 0.2063$. As demonstrated in Ref. \cite{amir_armesto}, 
this phenomenological dipole  reproduces quite well the HERA data on exclusive $\rho$ and $J/\Psi$ production. Moreover, the results from 
Refs. \cite{bruno1,bruno2} demonstrated that this model gives a good description of the recent LHC data on exclusive vector meson photoproduction 
in $pp$ and $pPb$ collisions.

The main assumptions in our approach for  inclusive and exclusive leading neutron processes are that i) the absorptive corrections can be represented by  
a ${\cal{K}}$ factor, which is assumed to be energy and $x_L$ independent and ii) the dipole - pion amplitude can be related to the dipole - proton one  
by a constant factor $R_q$. Both assumptions surely deserve more detailed studies (For  more discussions see Refs. \cite{nosLN,nosLN2}). However, the results 
presented in Refs. \cite{nosLN,nosLN2} demonstrated that these assumptions are supported by the available HERA data. In particular, in Ref. \cite{nosLN2} we 
have assumed $R_q = 2/3$, as expected from the additive quark model, and we have used the experimental HERA data on  
$\sigma (\gamma p \rightarrow \rho  \otimes \pi + n)$ \cite{rhoLN_HERA} to constrain the range of possible values of the $\cal{K}$ - factor. 
In the calculations of the exclusive vector meson production associated with a leading neutron in photon - induced interactions we  will also assume that 
$R_q = 2/3$ and we will use the  values of the  $\cal{K}$ - factor in the range determined in Ref. \cite{nosLN2}. As a consequence, the predictions 
to be presented in the next section  are parameter -- free. Therefore, the analysis of this process in hadronic colliders, which have  cross sections 
larger than those studied at HERA, will be very useful to test our approach and its underlying assumptions.  

\section{Results}
\label{res}

\begin{figure}[t]
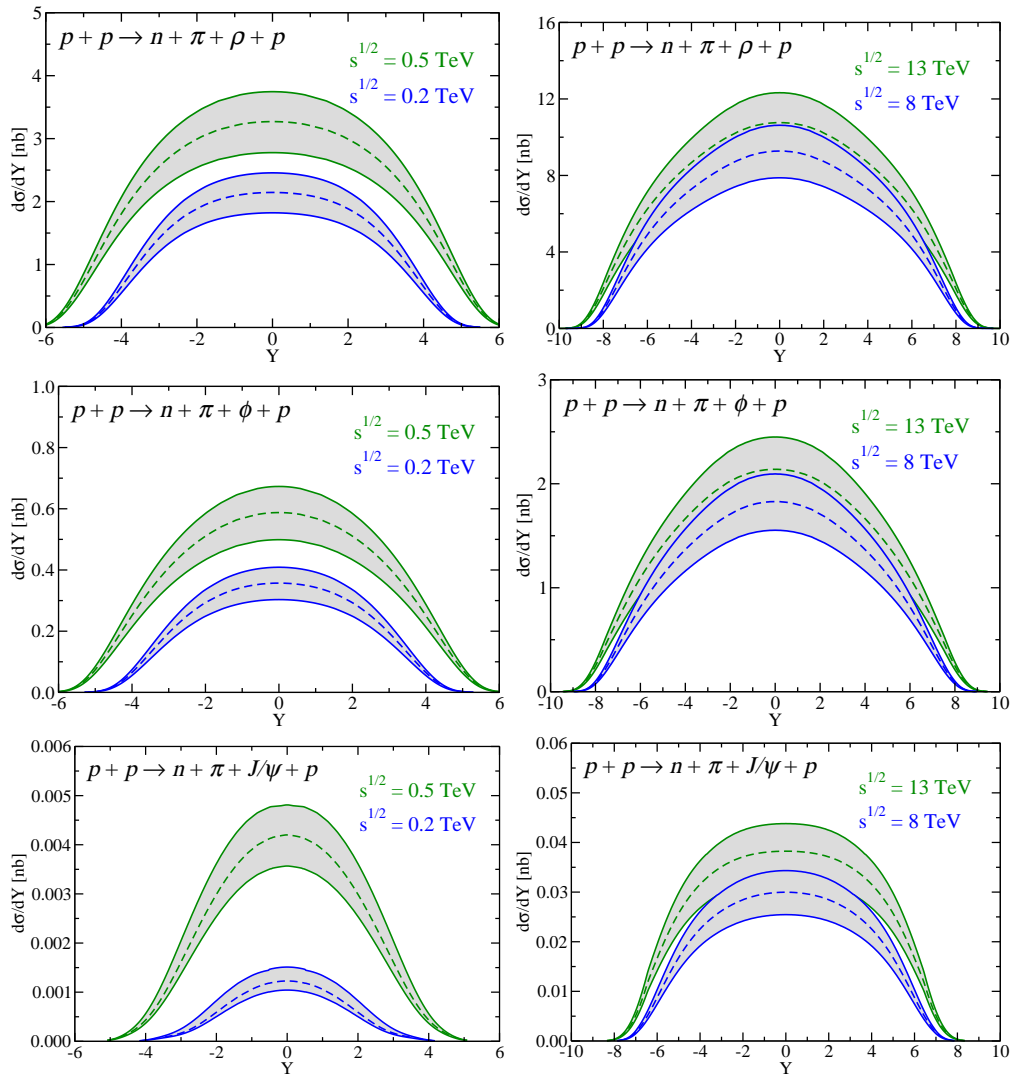

\begin{tabular}{cc}
{\psfig{figure=rho_pp_RHIC.eps,width=6.5cm}} & {\psfig{figure=rho_pp_LHC.eps,width=6.5cm}}  \\
{\psfig{figure=phi_pp_RHIC.eps,width=6.5cm}} & {\psfig{figure=phi_pp_LHC.eps,width=6.5cm}} \\
{\psfig{figure=jpsi_pp_RHIC.eps,width=6.5cm}} & {\psfig{figure=jpsi_pp_LHC.eps,width=6.5cm}}
\end{tabular}
\caption{Rapidity distribution for the exclusive $\rho$, $\phi$ and $J/\Psi$ production associated with a leading neutron in $\gamma p$ interactions at $pp$ collisions.  }
\label{fig2}
\end{figure}

In this section we present our predictions for the exclusive vector meson production associated with a leading neutron in photon - induced 
interactions considering $pp/pA$ collisions at RHIC and LHC energies. In particular, we will consider $pp$ collisions at $\sqrt{s} = 0.2, \, 0.5, \, 8$ 
and 13 TeV,  $pAu$ collisions at $\sqrt{s} = 0.2$ and 0.5 TeV and $pPb$ collisions at $\sqrt{s} = 5$ and 8.8 TeV. In order to estimate the total cross 
section $\gamma p \rightarrow V  \otimes \pi + n$, given by Eq. (\ref{crossgen}), we will assume that $p_T = \sqrt{|t|} <  0.2$ GeV, as implemented in 
the analysis of the H1 Collaboration \cite{rhoLN_HERA}. Moreover, in \cite{nosLN2} we have estimated three possible values for the $\cal{K}$ - factor 
considering the central value of the total cross section for  exclusive $\rho$ photoproduction with a leading neutron and its upper and lower bounds 
determined in Ref. \cite{rhoLN_HERA}. In what follows we will assume these same values, which are given by $({\cal{K}}_{min}, {\cal{K}}_{med}, 
{\cal{K}}_{max}) = (0.152, 0.179, 0.205)$. As a consequence, instead of a single curve for the rapidity distribution we will obtain a band. Similarly, we 
will derive a range of possible values for the total cross sections.

In Fig. \ref{fig2} we present our predictions for the rapidity distributions of the vector mesons produced in $pp$ collisions at  $\sqrt{s} = 0.2, \, 0.5, \, 8$ 
and 13 TeV. As expected from the symmetry of the initial state, the distributions are symmetric with respect to $Y = 0$. Moreover, the predictions for 
midrapidities $Y \approx 0$ increase with the energy and decrease for heavier vector mesons. Additionally, the growth with the energy is faster for  $J/\Psi$ 
production. This can be directly associated to the fact that charmonium production is dominated by color dipoles of small size.  On the other hand, the overlap 
functions $(\Psi^{V*}\Psi)$ of the lighter mesons ($\rho$ and $\phi$) peak at larger pair separations at a fixed photon virtuality \cite{kmw}. As a consequence, 
the impact of the non-linear effects is larger  for these states and they reduce the growth of the cross sections with the energy.

\begin{figure}[t]
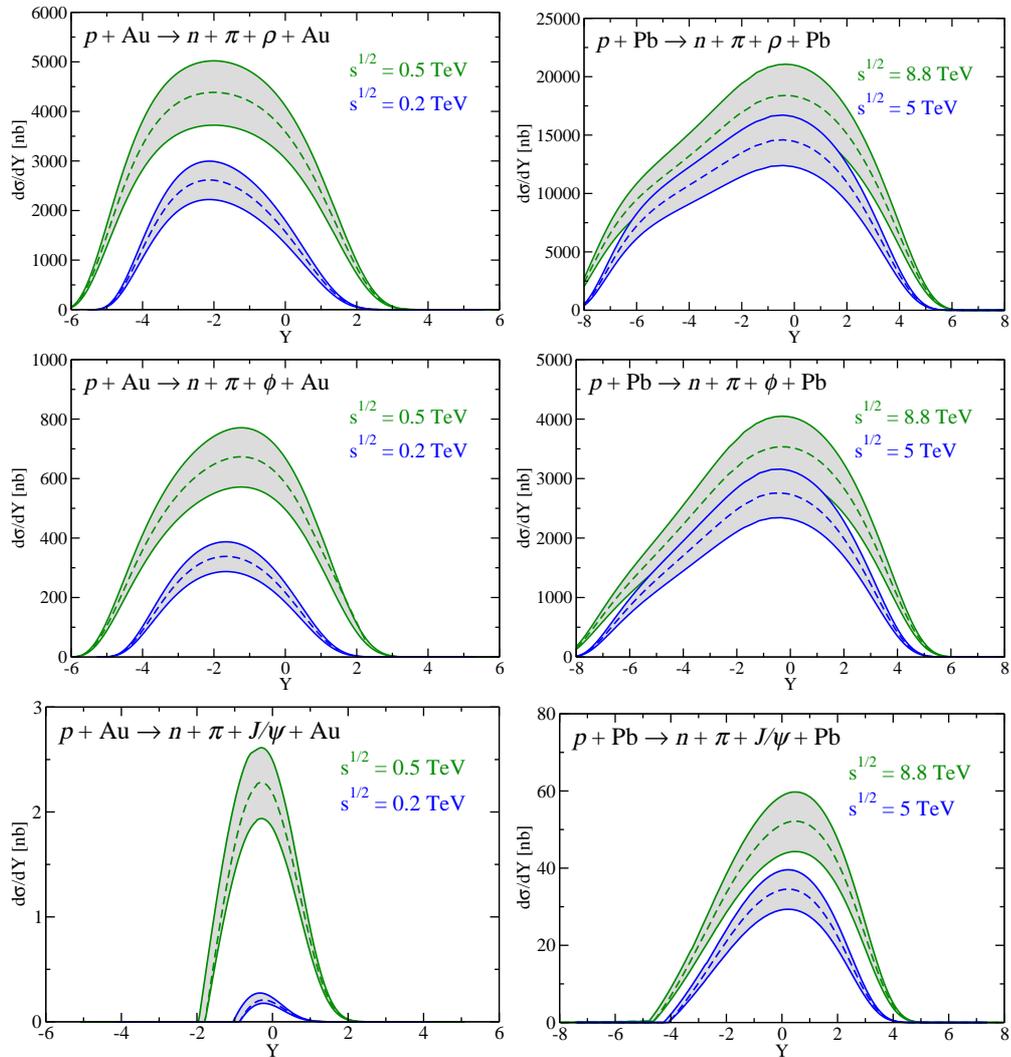

\begin{tabular}{cc}
{\psfig{figure=rho_pA_RHIC.eps,width=6.5cm}} & {\psfig{figure=rho_pA_LHC.eps,width=6.5cm}}  \\
{\psfig{figure=phi_pA_RHIC.eps,width=6.5cm}} & {\psfig{figure=phi_pA_LHC.eps,width=6.5cm}} \\
{\psfig{figure=jpsi_pA_RHIC.eps,width=6.5cm}} & {\psfig{figure=jpsi_pA_LHC.eps,width=6.5cm}}
\end{tabular}
\caption{Rapidity distribution for the exclusive $\rho$, $\phi$ and $J/\Psi$ production associated with a leading neutron in $\gamma p$ interactions at $pA$ collisions.  }
\label{fig3}
\end{figure}

In Fig. \ref{fig3}  we present our predictions for the rapidity distributions of the vector mesons produced in $pAu$ collisions at  $\sqrt{s} = 0.2$ and  
0.5  TeV as well as in $pPb$ collisions at  $\sqrt{s} = 5.0$ and  8.8 TeV. In this case we have asymmetric distributions, since the photon - induced 
interactions are dominated by photons emitted by the nucleus. This behaviour is directly associated to the fact that the photon flux of a nucleus is 
proportional  to $Z^2$. This enhancement has direct impact on the magnitude of the distributions, which  are amplified in comparison with the $pp$ one.

In Tables \ref{total_pp} and \ref{total_pA} we present our predictions for the total cross sections considering $\gamma p$ interactions in $pp$ and $pA$ 
collisions, respectively. As expected from the analysis of the rapidity distributions, the cross sections increase with the energy and decrease with the mass 
of the vector meson. Moreover, their magnitude is enhanced in $pA$ collisions in comparison with the predictions for  vector meson production in $pp$ ones.  
In the case of  $\rho$ production, we predict values of the order of $10^1$ ($10^4$) nb in $pp$ ($pAu$) collisions at RHIC energies. At LHC energies, we 
predict that the total cross section will be   $ \approx 10^2$ ($10^5$) nb in  $pp$ ($pPb$) collisions. In comparison with the photoproduction of vector mesons 
in $\gamma p$ interactions at hadronic collisions in processes without the presence of a leading neutron (See, e.g. Refs. \cite{vicmag_mesons1,vicmag_update,
Schafer,motyka_watt,bruno1,bruno2,glauber}),  our predictions are smaller by approximately two orders of magnitude, as expected from the experimental results 
obtained in $\gamma p$ collisions at HERA. However, it is important to emphasize that these events will be characterized by a very forward neutron, which can 
be used to tag the events and allow for the study the process. This possibility has been discussed e.g. in Refs. \cite{star,contreras}. In particular, in Ref. 
\cite{contreras} the author has analysed  the capabilities of the neutron Zero Degree Calorimeters detectors  to select  events with one leading forward neutron, 
which could be tagged and then studied using the central ALICE detector. Finally, considering the design luminosities for $pp$ and $pA$ collisions at LHC given 
by  ${\cal L}^{\mathrm{pp}} = 10^7$ mb$^{-1}$s$^{-1}$ and ${\cal L}^{\mathrm{pPb}} = 150$ mb$^{-1}$s$^{-1}$, respectively, and assuming a run time of $10^7 \, 
(10^6)$ s for collisions with protons (ions),  we predict that the $\rho$ production rates  will be of the order of $10^{10}$ ($10^7$) events per year. 
In the case of  $J/\Psi$ production, we predict $10^{7}$ ($10^4$) events per year. These results demonstrate that the experimental analysis of  vector meson 
production with a leading neutron in $\gamma p$ interactions at $pp$ and $pA$ collisions is, in principle, feasible at the LHC.

\section{Summary}
\label{conc}
Our understanding of the  hadron structure, the QCD dynamics and the description of inclusive and exclusive processes has  advanced with the successful operation 
of the DESY $ep$ collider HERA.  However, several questions remains without answer. As HERA has stopped to operate and the next generation of $ep$ collider is  
still to be constructed, the study of alternatives which could help to study the processes mentioned above are timely and necessary. One possibility is the use of 
the hadronic colliders to study  photon - induced interactions in a new kinematical range of $\gamma h$ center - of - mass energies. Recent results have demonstrated 
that the analysis of these processes is feasible at  RHIC and LHC, and that it is possible to use the resulting experimental data to investigate e.g.  the nuclear 
effects in the gluon distribution, the QCD dynamics at high energies and several other issues that still lack  a satisfactory description. This possibility  has 
stimulated the improvement of the theoretical description of these processes as well as the proposal of new forward detectors to be installed in the LHC \cite{ctpps,marek}.
One important issue  is the understanding of the leading neutron processes, which have strong implications in the forward physics at colliders and ultrahigh energy 
cosmic rays. Recently, we proposed a model to describe  inclusive and exclusive leading neutron processes in $ep$ collisions, which reproduce quite well the HERA data. 
However, several questions deserve more detailed studies and the comparison of these predictions, as well those from other models of leading neutron processes, 
with a larger and more precise set of experimental data is fundamental to improve our understanding of these processes. In this paper we proposed the analysis of 
vector meson production associated with a leading neutron in $\gamma p$ interactions at $pp$ and $pA$ collisions  as an alternative to study  leading neutron processes.   
Considering that all the elements of our model have been constrained by the HERA data, we have presented  parameter -- free predictions for $\rho$, $\phi$ and 
$J/\Psi$ production with a leading neutron at RHIC and LHC energies. We predicted large values for the total cross sections and event rates, which implies that the 
experimental analysis of this process is, in principle, feasible. We expect that our results motivate future experimental analysis at RHIC and LHC colliders, 
which undoubtedly will allow to constrain the description of the leading neutron processes.

\begin{table}[t]
\centering
\begin{tabular}{||c|c|c|c|c|c||}
\hline
\hline
$\sigma(V)$ [nb] &                             & $\sqrt{s} = 0.2$ TeV     & $\sqrt{s} = 0.5$ TeV     & $\sqrt{s} = 8.0 $  TeV  &  $\sqrt{s} = 13.0$ TeV \\ \hline \hline
 \phantom{$\sigma(\rho)$ [nb]}  &  $K_{min}$ & 12.17  & 22.06  & 90.12  & 110.51 \\
 $\rho$             & $K_{med}$ & 14.34  & 25.98  & 106.12 & 130.14 \\
 \phantom{$\sigma(\rho)$ [nb]}   & $K_{max}$ & 16.42  & 29.75  & 121.54 & 149.04 \\ \hline \hline
 \phantom{$\sigma(\phi)$ [nb]}   & $K_{min}$ & 1.83   & 3.58   & 16.67  & 20.73  \\
 $\phi$             & $K_{med}$ & 2.15   & 4.21   & 19.63  & 24.42  \\
 \phantom{$\sigma(\phi)$ [nb]}   & $K_{max}$ & 2.46   & 4.83   & 22.48  & 27.96  \\ \hline \hline
  \phantom{$\sigma(J/\psi)$ [nb]}& $K_{min}$ & 0.0042 & 0.019 & 0.25 & 0.35 \\
 $J/\psi$           & $K_{med}$ & 0.0049 & 0.022 & 0.30 & 0.42 \\
 \phantom{$\sigma(J/\psi)$ [nb]} & $K_{max}$ & 0.0064 & 0.026 & 0.34 & 0.48 \\ \hline
 \hline
\end{tabular}
\caption{Total cross sections for the exclusive $\rho$, $\phi$ and $J/\Psi$ production associated with a leading neutron in $pp$ collisions considering 
different center - of - mass energies and distinct values for the magnitude of the absorptive corrections, described by the $\cal{K}$ - factor.}
\label{total_pp}
\end{table}

\begin{table}[t]
\centering
\begin{tabular}{||c|c|c|c|c|c||}
\hline
\hline
 $\sigma(V)$ [nb] &                             & $\sqrt{s} = 0.2$ TeV     & $\sqrt{s} = 0.5$ TeV     & $\sqrt{s} = 5.0$ TeV    &  $\sqrt{s} = 8.8$ TeV \\ \hline \hline
 \phantom{$\sigma(\rho)$ [nb]} &    $K_{min}$ & 9176.88 & 20819.90 & 102785.00  &  139110.00 \\
 $\rho$ &              $K_{med}$ & 10807.00 & 24518.30 & 121043.00  &  163821.00 \\
 \phantom{$\sigma(\rho)$ [nb]}  &  $K_{max}$ & 12376.70 & 28079.50 & 138625.00  &  187616.00 \\ \hline \hline
 \phantom{$\sigma(\phi)$ [nb]}  &  $K_{min}$ & 1090.55 & 2863.67 & 17326.20 &  24154.90 \\
 $\phi$             & $K_{med}$ & 1278.41 & 3386.65 & 20403.80 &  28445.60 \\
 \phantom{$\sigma(\phi)$ [nb]}   & $K_{max}$ & 1470.81 & 3862.20 & 23367.50 &  32577.30 \\ \hline \hline
  \phantom{$\sigma(J/\psi)$ [nb]} & $K_{min}$ & 0.19 & 3.94 & 135.53 &  234.50 \\
 $J/\psi$            & $K_{med}$ & 0.23 & 4.65 & 159.61 &  276.16 \\
 \phantom{$\sigma(J/\psi)$ [nb]} & $K_{max}$ & 0.32 & 5.65 & 184.04 &  317.05 \\ \hline \hline
\end{tabular}
\caption{Total cross sections for the exclusive $\rho$, $\phi$ and $J/\Psi$ production associated with a leading neutron in $pAu$ collisions at $\sqrt{s} = 0.2$ 
and 0.5 TeV and $pPb$ collisions at at $\sqrt{s} = 5.0$ and 8.8 TeV considering different  values for the magnitude of the absorptive corrections, described by 
the $\cal{K}$ - factor.}
\label{total_pA}
\end{table}

\begin{acknowledgments}
VPG would like to thank G. Contreras, S. Klein and D. Tapia Takaki by useful discussions. 
This work was  partially financed by the Brazilian funding agencies CNPq, CAPES, FAPERGS and FAPESP.

\end{acknowledgments}

\hspace{1.0cm}

\end{document}